\documentclass[apj]{emulateapj}
\newcommand{\farc}{\hbox{$.\!\!^{\prime\prime}$}} 
\usepackage{graphicx}
\usepackage{txfonts}
\usepackage[stable]{footmisc}

\usepackage{natbib}
\bibpunct[, ]{(}{)}{;}{a}{}{,}

\newcommand{\hb}{H$\beta$} 
\newcommand{\ha}{H$\alpha$}

\newcommand{\oii}{[\ion{O}{2}]} 
\newcommand{\oiii}{[\ion{O}{3}]}

\slugcomment{ApJ in press}
\begin{document}

\shorttitle{TOUGH5: X-shooter redshifts}
\shortauthors{Thomas Kr\"{u}hler et al., - TOUGH V: X-shooter redshifts}


\title{The optically unbiased GRB host (TOUGH) survey. \\V. VLT/X-shooter emission-line redshifts for \textit{Swift} GRBs at $z\sim2$\altaffilmark{\dag}}

\altaffiltext{\dag}{Based on observations collected at the European Southern Observatory, Paranal, Chile, Program IDs: 177.A-0591, 080.A-0825, 084.A-0303, 086.B-0954, 087.B-0737, 088.A-0644.}

\author{Thomas~Kr\"{u}hler\altaffilmark{1},\
Daniele~Malesani\altaffilmark{1},\
Bo~Milvang-Jensen\altaffilmark{1},\
Johan~P.~U. Fynbo\altaffilmark{1},\
Jens~Hjorth\altaffilmark{1},\
P\'all~Jakobsson\altaffilmark{2},\
Andrew~J.~Levan\altaffilmark{3},\
Martin~Sparre\altaffilmark{1},\
Nial~R.~Tanvir\altaffilmark{4},
Darach~J. Watson\altaffilmark{1}
}

\altaffiltext{1}{Dark Cosmology Centre, Niels Bohr Institute, University of Copenhagen, Juliane Maries Vej 30, 2100 Copenhagen, Denmark.}
\altaffiltext{2}{Centre for Astrophysics and Cosmology, Science Institute,
University of Iceland, Dunhagi 5, 107 Reykjavik, Iceland}
\altaffiltext{3}{Department of Physics, University of Warwick, Coventry CV4 7AL, UK}
\altaffiltext{4}{Department of Physics and Astronomy, University of
Leicester, University Road, Leicester, LE1 7RH, UK}

\begin{abstract}

We present simultaneous optical and near-infrared (NIR) spectroscopy of 19 \textit{Swift} GRB host galaxies with VLT/X-shooter with the aim 
of measuring their redshifts. Galaxies were selected from The Optically Unbiased GRB Host (TOUGH) survey (15 of the 19 galaxies) or because they hosted GRBs without a bright optical afterglow. Here, we provide emission-line redshifts for 13 of the observed galaxies with brightnesses between $\rm{F606W}>27\,\rm{mag}$ and $R=22.9\,\rm{mag}$ (median $\tilde{R}=24.6\,\rm{mag}$). The median redshift is $\tilde{z}=2.1$ for all, and $\tilde{z}=2.3$ for the TOUGH hosts. Our new data significantly improve the redshift completeness of the TOUGH survey, which now stands at 77\% (53 out of 69 GRBs). They furthermore provide accurate redshifts for nine prototype-dark GRBs (e.g., GRB 071021 at $z=2.452$ and GRB 080207 at $z=2.086$), which are exemplary of GRBs where redshifts are challenging to obtain via afterglow spectroscopy. This establishes X-shooter spectroscopy as an efficient tool for redshift determination of faint, star-forming, high-redshift galaxies such as GRB hosts. It is hence a further step towards removing the bias in GRB samples that is caused by optically-dark events, and provides the basis for a better understanding of the conditions in which GRBs form. The distribution of column densities as measured from X-ray data ($N_{\rm{H, X}}$), for example, is closely related to the darkness of the afterglow and skewed towards low $N_{\rm{H, X}}$ values in samples that are dominated by bursts with bright optical afterglows. 
 
\end{abstract}

%

\section{Introduction}

Our understanding of long $\gamma$-ray bursts \citep[GRBs, e.g.,][]{2004RvMP...76.1143P, 2009ARA&A..47..567G}, their multi-wavelength afterglows, and star-forming host galaxies fundamentally relies on the information about the distance scale to these events. The redshift of a GRB is most robustly obtained via optical/near-infrared (NIR) follow-up observations: afterglow spectroscopy yields a precise redshift measurement through the signature of the host's interstellar medium (ISM) from fine-structure transitions 
\citep[][]{2006ApJ...648...95P, 2007A&A...468...83V} or hydrogen and metal lines \citep[][]{2009ApJS..185..526F}. Also multi-band photometry provides a rough distance scale through the identification of the strong Ly$\alpha$ and Ly-limit breaks \citep[][]{2011A&A...526A.153K}. 

Redshifts\footnote{\texttt{http://www.mpe.mpg.de/$\sim$jcg/grbgen.html}}$^{,}$\footnote{ \texttt{http://www.raunvis.hi.is/$\sim$pja/GRBsample.html}} and host data\footnote{\texttt{http://www.grbhosts.org}} 
are publicly available for a large sample of GRBs. The prerequisite of an optical afterglow for redshift measurements, however, causes substantial selection biases due to the phenomenon of optically-dark GRBs \citep[][]{1998ApJ...493L..27G, 2001ApJ...562..654D}. A high redshift and significant column densities of dust along the GRB's sight line efficiently suppress afterglow emission in the UV/optical wavelength bands, where most of the follow-up observations and redshift measurements are performed \citep[e.g.,][]{2004ApJ...617L..21J}. 

This leads to strong selection effects, which arguably play a crucial role when using GRBs as tools, in the distribution of metallicities of GRB damped Ly$\alpha$ absorbers \citep[DLAs,][]{2009ApJ...691L..27P}, the dust characteristics of GRB sight-lines \citep[][]{2011A&A...526A..30G, 2011arXiv1102.1469Z} as well as the metallicity, luminosity, mass-distribution and Ly$\alpha$-emission of long GRB hosts \citep[][]{2011arXiv1108.0674K, 2012arXiv1205.3779M}.

{A different route to redshift measurements for GRBs without an optical afterglow exists through an association between burst and its host galaxy and subsequent galaxy spectroscopy. This requires GRB positions with an uncertainty of at most few arcseconds. In the absence of optical/NIR afterglows, those positions were previously obtained via radio/sub-mm or X-ray observations \citep[e.g.,][]{2001ApJ...562..654D, 2002ApJ...577..680P, 2005ApJ...629...45J, 2007A&A...475..101C, 2007ApJ...669.1098R, 2007ApJ...660..504B, 2011GCN..12190...1Z}. The X-Ray Telescope \citep[XRT, ][]{2005SSRv..120..165B} onboard \textit{Swift} \citep{2004ApJ...611.1005G} provides these positions (accuracy of \citealp[$\lesssim2\farcs0$,][]{2007A&A...476.1401G, 2007AJ....133.1027B}) for the first time with large number statistics ($\sim 100$ events per year), facilitating the build-up of large, well-defined samples of GRBs, afterglows and hosts with minimal optical biases \citep{2012arXiv1205.3162H}}. 

GRB host spectroscopy is however mostly performed with optical spectrographs \citep[][]{2009ApJ...691..182S, 2010AJ....139..694L}. This gives rise to a redshift desert ($1.4\lesssim{z}\lesssim 2.5$) where the strongest recombination and forbidden lines are outside the optical wavelength range. Spectroscopic studies of GRB hosts were thus limited to the low-redshift end of the GRB redshift distribution until NIR spectroscopy became feasible \citep[][]{TK2012, 2012MNRAS.419.3039C}.

Here we present simultaneous optical/NIR spectroscopy with VLT/X-shooter for 19 \textit{Swift} GRB hosts galaxies
with the aim of increasing the redshift completeness of The Optically Unbiased GRB Host (TOUGH\footnote{\texttt{http://www.dark-cosmology.dk/TOUGH}}, \citealt{2012arXiv1205.3162H, Daniele2012, 2012ApJ...752...62J, 2012arXiv1205.3779M, 2012arXiv1205.4239M} and this work) survey. Our observations provide accurate redshifts for 13 \textit{Swift} GRBs, out of which nine GRBs are characterized as dark, i.e., with suppressed afterglow flux in the optical wavelength range as compared to the X-ray regime. We use our new redshift determinations to illustrate the bias in previous GRB samples primarily based on afterglow spectroscopy.

\section{Observations and Data Reduction}

\subsection{Sample Selection}

The GRB hosts in this work have been selected from two sources (Table~\ref{tab:obs}).
The bulk of the sample was taken from the TOUGH survey, while the
remaining hosts were specifically chosen because of the dark nature of their afterglows. Dark GRBs were selected because they were localized to sub-arcsecond accuracy, either from \textit{Chandra} X-ray or NIR ground-based afterglow observations, yielding reliable host associations. 

Furthermore, we required that the hosts had existing imaging for a precise slit-alignment. Pre-imaging has been obtained with VLT FORS2 or ISAAC \citep{Daniele2012} 
and HST ACS or WFC3 \citep{Dan2012}. 
Finding charts centered on the respective hosts are shown in Figures~\ref{fct} and \ref{fc}. The four galaxies not part of TOUGH (GRBs 071021, 080207, 090113, 090407) have brightnesses\footnote{All magnitudes in this work are given in the Vega system.} between $\rm{F606W}>27\,\rm{mag}$ and $R\sim24.5\,\rm{mag}$ and $K\sim21\,\rm{mag}$ to $K\sim19.5\,\rm{mag}$, making them comparatively red and NIR bright. The brightness of the TOUGH hosts in this paper is in the range $26.5\,\rm{mag}\gtrsim\textit{R}\gtrsim22.9\,\rm{mag}$ ($\textit{K}\gtrsim20.0\,\rm{mag}$).

{\begin{figure}
\centering
\includegraphics[angle=0, width=0.7\columnwidth]{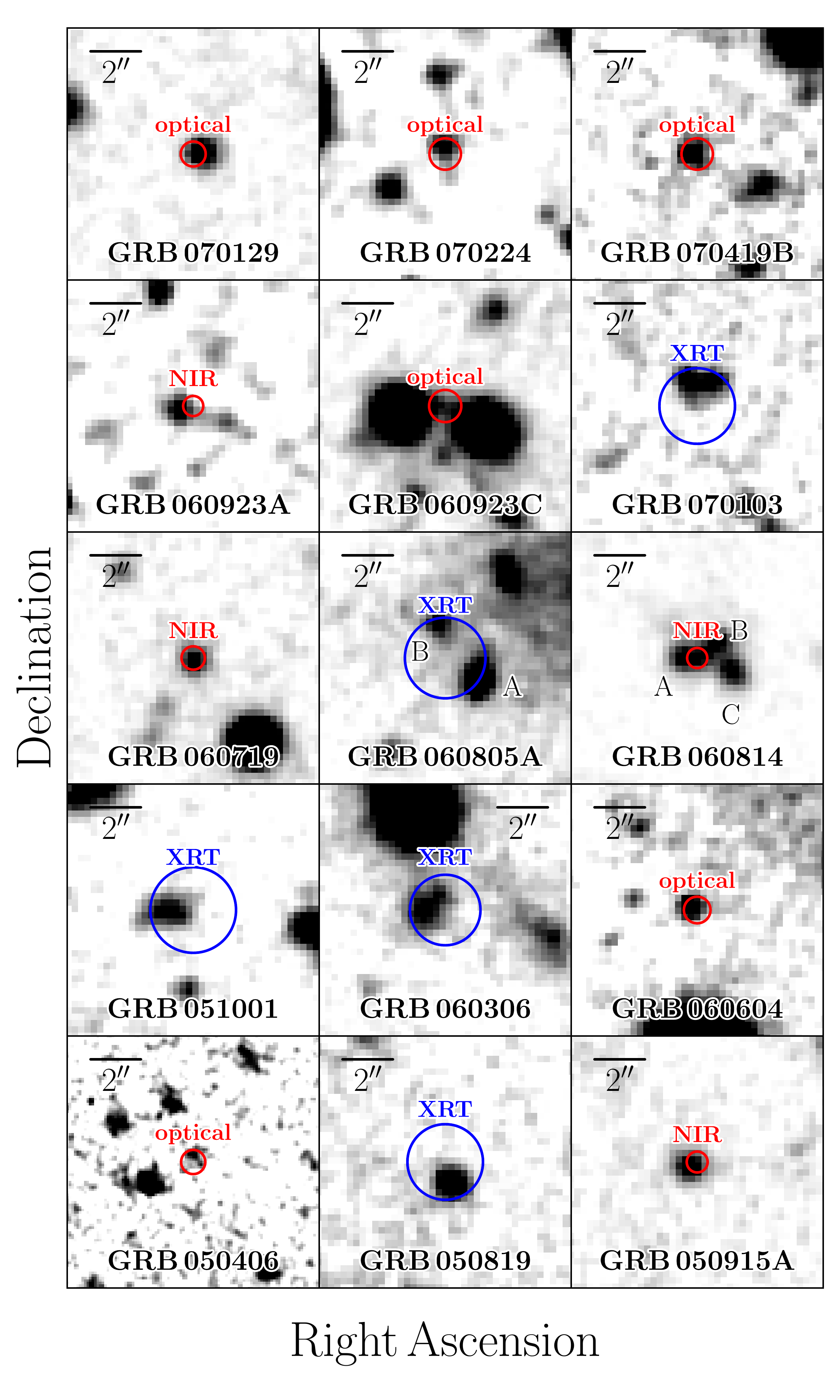}
\caption{VLT/FORS $R$-band finding charts ($10\arcsec\,\times\,10\arcsec$) for our 15 TOUGH GRBs where North is up and East is left in all panels. Afterglow positions (ground-based optical/NIR or \textit{Swift}/XRT X-ray) are indicated by red and blue circles. Photometry, positions and errors are described in detail in \citet{Daniele2012}. Images are smoothed and the radii of optical/NIR error circles are indicative to enhance clarity. In cases where more than one galaxy is detected in the vicinity of the respective error-circles, the most likely host is labeled with A.}
\label{fct}
\end{figure}
}
\begin{figure}
\includegraphics[angle=0, width=0.99\columnwidth]{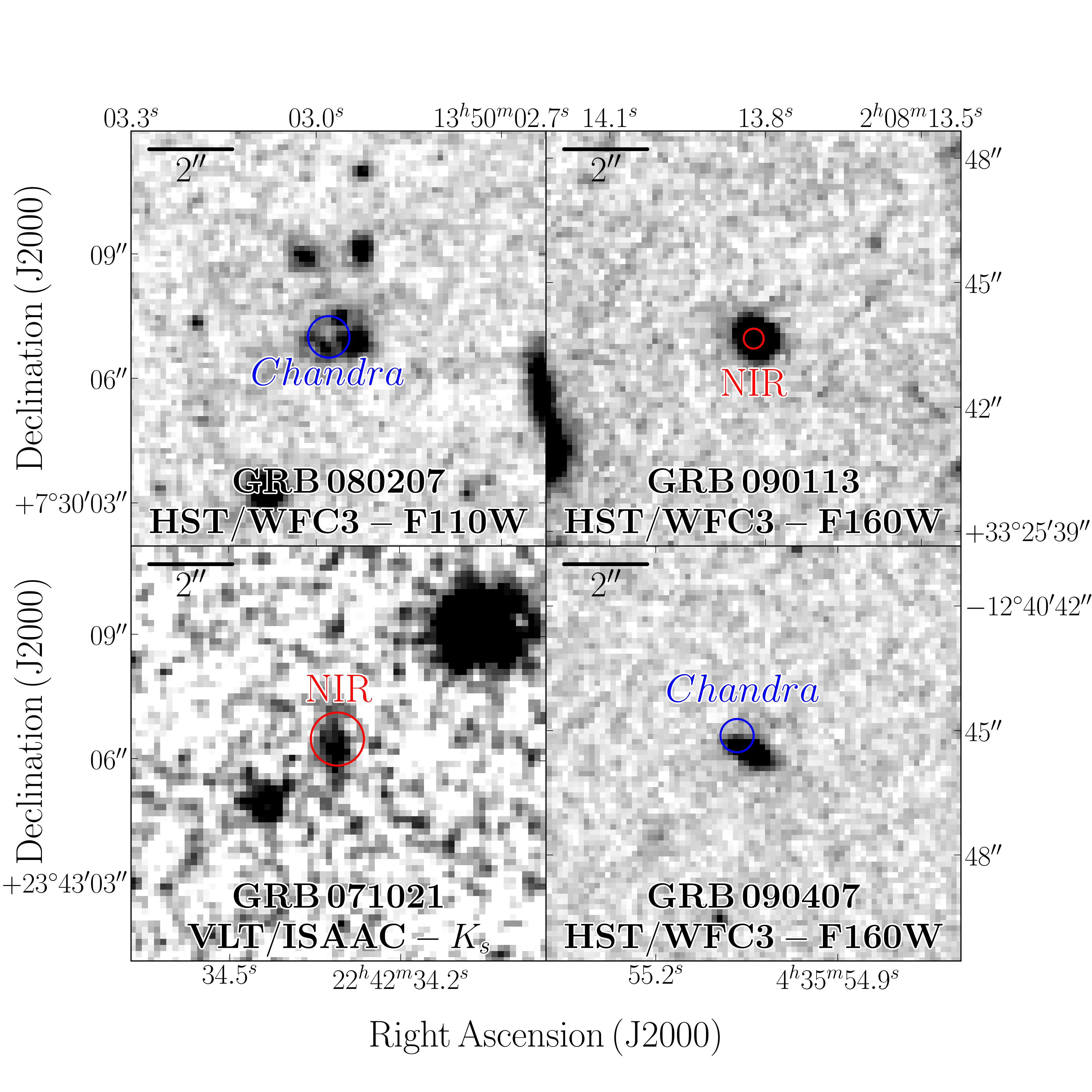}
\caption{Finding charts ($10\arcsec\,\times\,10\arcsec$) for the four non-TOUGH GRBs 071021, 080207, 090113, and 090407 where North is up and East is left in all panels. Afterglow positions (ground-based NIR or \textit{Chandra} X-ray) are indicated by red and blue circles. Photometry, positions and errors are described in detail in \citet{Dan2012}.}
\label{fc}
\end{figure}

\subsection{X-shooter Observations}

\begin{deluxetable*}{ccccccccccc}
\tabletypesize{\scriptsize}
\tablecaption{Details of GRB Host Selection, Observations and Redshifts
\label{tab:obs}}
\tablecolumns{11} 
\tablehead{
	\colhead {GRB/Host}& 
	\colhead {$\beta_{\rm{OX}}^{(a)}$} & 
	\colhead {Brightness} & 
	\colhead {$p_{\rm{chance}}^{(b)}$} & 
	\multicolumn{2}{c}{Exposure time (s)} & 
	\colhead {Airmass} & 
	\colhead {Seeing} & 
	\colhead {Selection$^{(c,d)}$} &
	\colhead {Emission Lines$^{(h)}$} &
	\colhead {Redshift} \\
	\colhead {} & 
	\colhead {} &
	\colhead {$R$-band (mag)} & 
	\colhead {} &
	\colhead {UVB/VIS} & 
	\colhead {NIR}  &  
	\colhead {} & 
	\colhead {(\arcsec)} & 
	\colhead {} & 
   	\colhead {} & 
	\colhead {} } 
\startdata
050406$^{(c)}$& 1.0 & 26.6  & 0.02 &{$3\times1800$}&$6\times900$& 1.3 & 0\farcs{8} & TOUGH & \nodata & \nodata \\
050819$^{(c)}$& $<0.9$ & 24.1 & 0.03 &{$2\times900$}&$4\times450$& 1.7 & 0\farcs{7} & TOUGH & (\oii,~\hb), \oiii& $2.5043\pm0.0007$  \\
050915A$^{(c,d)}$& $<0.4$ & 24.6  & 0.004 &{$6\times1200$}& $12\times600$  & 1.3 & 0\farcs{8} &  TOUGH & \oii, \hb, (\oiii) & $2.5273\pm0.0003$ \\
051001$^{(c,d)}$& $<0.6$ & 24.4  & 0.05 &{$2\times1800$}& $4\times900$  & 1.1 & 0\farcs{7} &  TOUGH & \hb, \oiii, (\ha) & $2.4296\pm0.0003$ \\
060306$^{(c,d)}$& $<0.5$ &  24.1 & 0.03 &{$2\times900/2\times938$}& $2\times2\times490$ &  1.2 & 1\farcs{9} &  TOUGH & \nodata & \nodata \\
060604$^{(c)}$& 0.8 & 25.5  & 0.01 &{$4\times1800$}& $8\times900$ &  1.1 & 0\farcs{8} &  TOUGH & \hb, \oiii, \ha & $2.1357\pm0.0002$ \\
060719$^{(c,d)}$& $<-0.1$ & 24.6 & 0.005 &{$4\times1500$}& $12\times500$ & 1.2 & 0\farcs{7} & TOUGH & (\oii), \ha & $1.5320\pm0.0004$ \\
060805A$^{(c,d)}$& $<1.1$& 23.5 & 0.02  &{$4\times1200$}& $12\times420$ & 1.4 & 2\farcs{0} & TOUGH & \nodata & \nodata \\
060814$^{(c,d)}$& $<-0.1$ & 22.9 & 0.002 &{$2\times1800$}& $2\times3\times600$ &  1.5 & 0\farcs{8} & TOUGH & \oii, \oiii, \ha & $1.9229\pm0.0004$ \\
060923A$^{(c,d,e)}$& $<0.1$ & 26.1  & 0.02 &{$6\times1200$}& $12\times600$ &  1.3 & 1\farcs{0} & TOUGH & \nodata & \nodata \\
060923C$^{(c,d)}$& $<0.3$ & 25.5  & 0.02 &{$2\times730/2\times680$}& $6\times270$ &  1.3 & 0\farcs{7} & TOUGH & \nodata & \nodata \\
070103$^{(c,d)}$& $<0.5$ & 24.2  & 0.04 &{$2\times900$}& $4\times450$ &   1.7 & 0\farcs{8} & TOUGH & (\hb), \oiii & $2.6208\pm0.0007$ \\
070129$^{(c,d)}$& $0.6$  & 24.4  & 0.004 &{$4\times900$}& $4\times3\times300$ & 1.7 & 1\farcs{4} &  TOUGH & \oii, \oiii, \ha  & $2.3384\pm0.0003$ \\
070224$^{(c)}$& 0.9 & 26.0  & 0.15 &{$6\times1200$}& $12\times600$ &   1.1 & 0\farcs{5} & TOUGH & \nodata & \nodata \\
070419B$^{(c,d)}$& 0.3 & 25.2 & 0.01 &{$6\times1200$}& $12\times600$ &   1.1 & 1\farcs{1} &  TOUGH & \oiii, \ha & $1.9588\pm0.0003$ \\
\hline
071021$^{(f)}$& $<0.4$ & $V =25.3^{(f)}$  & 0.007 &{$2\times1200$}& $4\times600$ &   1.6 & 0\farcs{9} &  dark & \hb, \oiii, \ha & $2.4520\pm0.0004$ \\
080207$^{(f)}$& $<0.3$ & $25.8^{(c,g)}$ & 0.01 &{$6\times1200$}& $4\times3\times600$ & 1.2 & 0\farcs{9} &  dark & \oiii, \ha & $2.0858\pm0.0003$ \\
090113$^{(f)}$& $<0.3$ & $V=24.6^{(f)}$ & 0.005 &{$2\times1800$}& $4\times900$ &   2.0 & 0\farcs{7} &  dark & \oii, \ha & $1.7493\pm0.0002$ \\
090407$^{(f)}$& $<0.4$ & F606W~$>27^{(f)}$ & 0.02 &{$2\times1800$}& $4\times900$ & 1.0 & 0\farcs{6} &  dark & \oii, (\hb), \ha & $1.4485\pm0.0006$
\enddata
\tablecomments{$^{(a)}$ Optical-to-X-ray spectral index from \citet{2009ApJS..185..526F} for TOUGH events. Afterglows with $\beta_{\rm{OX}} < 0.5$ are defined as dark by \citet{2004ApJ...617L..21J}. $^{(b)}$ Estimated chance probability of finding an unrelated galaxy along the line of sight to the GRB. $^{(c)}$ The TOUGH sample and photometry is described in \citet{2012arXiv1205.3162H} and \citet{Daniele2012}. $^{(d)}$ TOUGH VLT/FORS spectroscopy in \citet{2012ApJ...752...62J}. $^{(e)}$ \citet{2008MNRAS.388.1743T}. $^{(f)}$ \citet{Dan2012}. $^{(g)}$ \citet{2012arXiv1202.1434R}. $^{(h)}$ Line transitions in the X-shooter spectrum. Values in round parentheses denote tentative detections, i.e., significant below 3$\sigma$ or strongly affected by sky-line contamination. The horizontal line separates the objects in the TOUGH sample from four extra systems that
were observed within our program.}
\end{deluxetable*}

We observed the hosts spectroscopically with the VLT equipped with X-shooter \citep{2011arXiv1110.1944V}. 
X-shooter is a medium resolution, cross-dispersed, echelle spectrograph capable of obtaining simultaneous spectra from $3000$ to $25\,000\,\AA$ in three arms (ultra-violet/blue (UVB), visual (VIS) and NIR arm). 

Our hosts were observed in nodding mode, with effective integration times ranging from 0.4 to 2.0 hr. Slit widths of the different arms were typically chosen to be 1\farcs{0} (UVB) and 0\farcs{9} (VIS and NIR). This results in approximate resolving powers $\lambda/\Delta\lambda\approx\,5100,\,8800,\,5300$ for the UVB, VIS, and NIR arm, respectively. 
Details about our X-shooter observations can be found in Table~\ref{tab:obs}.

\subsection{Data Reduction}

X-shooter data were reduced with the ESO/X-shooter pipeline v.~1.3.7 \citep{2006SPIE.6269E..80G}. The wavelength solution was obtained against arc-lamp frames, leaving residuals with an RMS of better than 0.3 pixels in all cases ($\approx7.5\,\rm{km\,s}^{-1}$ at $1.2\,\mu$m). A cross-calibration of all individual NIR spectra using sky-lines provided consistent results. Robust absolute flux measurements in the broad wavelength range of X-shooter are challenging. They are most reliably performed by scaling the spectral continuum to broad-band imaging within a similar wavelength interval. As for most of our targets both the necessary imaging as well as the detection of the host continuum are absent, we defer any analysis beyond line identification and redshift determination to forthcoming works.

\section{Results and Discussion}

\subsection{Selection Effects}
\label{seleff}

In principle, X-shooter is capable of measuring redshifts via strong recombination or forbidden lines up to $z\sim5$, which is when \oii($\lambda\lambda\,3726, 3729)$~is redshifted out of the NIR range. However, several selection effects play an important role in the redshift distribution of our sample. 

Firstly, the target selection is primarily based on host imaging within the TOUGH survey, yielding brightnesses between $26.5\,\rm{mag} \gtrsim \textit{R} \gtrsim 23\,\rm{mag}$, which biases the redshift distribution towards low-redshift events. {This bias is evident, for example, from the distribution of host magnitudes for which a redshift is available via afterglow spectroscopy (see e.g., Fig.~11 in \citealt{2012arXiv1205.3162H}). GRB hosts at $z > 3$ are typically fainter than the sensitivity limit of the TOUGH survey ($R \sim 27\,\rm{mag}$). Similarly, the targeted X-shooter hosts of this work are on average fainter and at higher redshifts than, for example, the hosts of the FORS campaign \citep{2012ApJ...752...62J}. Overall, the TOUGH survey shows an obvious trend of a decreasing median apparent brightness with increasing redshift for GRB hosts (see e.g., Fig.~8 in \citealt{2012arXiv1205.3162H}).}

Secondly, most of the hosts were already observed with optical spectrographs \citep{2012ApJ...752...62J}. Only if the optical spectroscopy did not reveal emission lines were the hosts further targeted with X-shooter. 
As a result of these selection criteria we expect our galaxies to be located in or above the redshift desert ($1.4< z \lesssim 3$).  In addition, there are further redshift ranges where emission lines are located in regions of low sky transparency making redshift measurements challenging observationally for example between $2.7\lesssim{z} \lesssim2.9$. Above $z=2.9$, all lines except \oii~are located in the $K$-band, where the thermal background of X-shooter is high.

\subsection{Host Associations}

GRB redshifts from their putative host galaxies always rely on the probabilistic association between galaxy and GRB event. Typical probabilities of associating an unrelated galaxy with the GRB ($p_{\rm{chance}}$) being in the 1-5\% range for 2\farc{0}-sized error-boxes and galaxies with brightnesses of around $R \sim 25\,\rm{mag}$ \citep[][]{2009AJ....138.1690P}. 

For 14 events in our sample, the association between GRBs and galaxy is unambiguous because of accurate afterglow positions from optical/NIR (typical accuracy $\lesssim0\farc3$) or dedicated \textit{Chandra} X-ray (typical accuracy $\sim0\farc5$) imaging (Figures~\ref{fct} and \ref{fc}). Only in five cases does the association rely on a \textit{Swift}/XRT position (typical accuracy $\sim1\farc5$).

{Using the magnitudes from the TOUGH survey and \citet{Dan2012}, we derive probabilities of associating an unrelated galaxy along the line of sight to the GRB (see Table~\ref{tab:obs}) following \citet{2002AJ....123.1111B}. These chance probabilities are $p_{\rm{chance}}\lesssim0.15$ for all our events, or $p_{\rm{chance}}\lesssim0.05$ for those with redshifts. We caution, that these values are strictly valid only for single events, and if the individual GRB sight-line probes a random line of sight through the Universe \citep[see e.g.,][for a detailed discussion]{2002AJ....123.1111B, 2009AJ....138.1690P}. Further support to the host assignments is added in the case where redshift determinations were successful: our redshift measurement depends on nebular and forbidden lines and thus star-formation, which is closely related to GRB formation. Together with the small chance probabilities, we hence consider the physical association between GRB and galaxy, and thus redshift, robust.}

\subsection{Emission Lines and Redshifts}

\begin{figure*}
\includegraphics[angle=0, width=1.98\columnwidth]{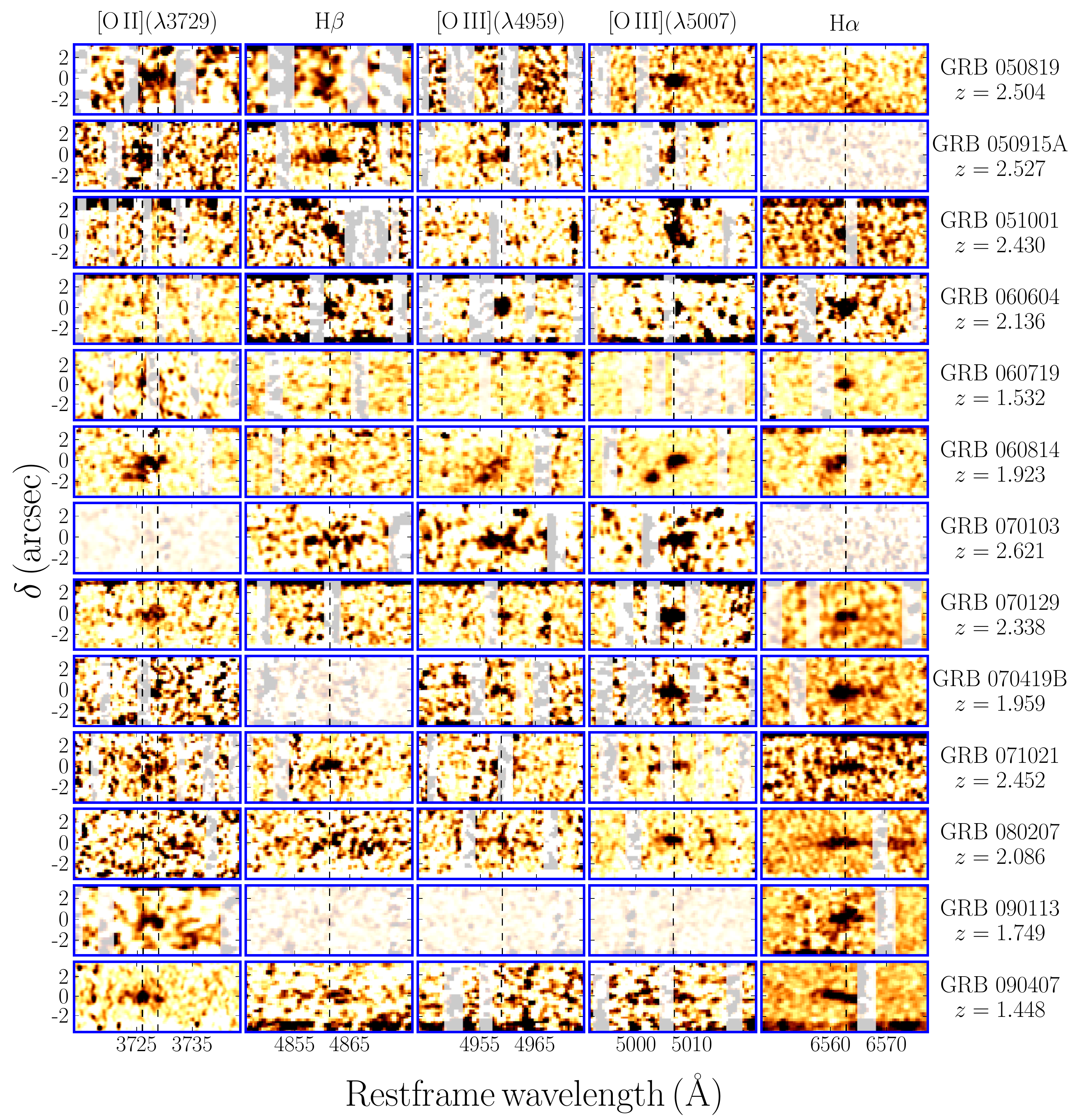}
\caption{Two-dimensional cutouts of 13 X-shooter spectra used for redshift determination. Each column is centered at the rest-frame wavelength of a strong emission line, marked by vertical dashed lines. Regions of sky-lines and negligible sky or X-shooter transmission are indicated with grey shading.}
\label{lines}
\end{figure*}

Our X-shooter spectroscopy provides accurate redshifts for 13 of the 19 targeted galaxies. Figure~\ref{lines} summarizes the detection of different recombination or forbidden lines and Table~\ref{tab:obs} shows the respective line transitions and redshifts. For eleven galaxies, two or more emission lines are detected at high significance yielding a robust and unique redshift. For two host galaxies (GRBs 050819 and 060719) only one line is detected at $\rm{S/N}>3$, and the line identification is thus not immediately clear. These are discussed in Sections \ref{050819} and \ref{060719}. 

\begin{figure}
\includegraphics[angle=0, width=.98\columnwidth]{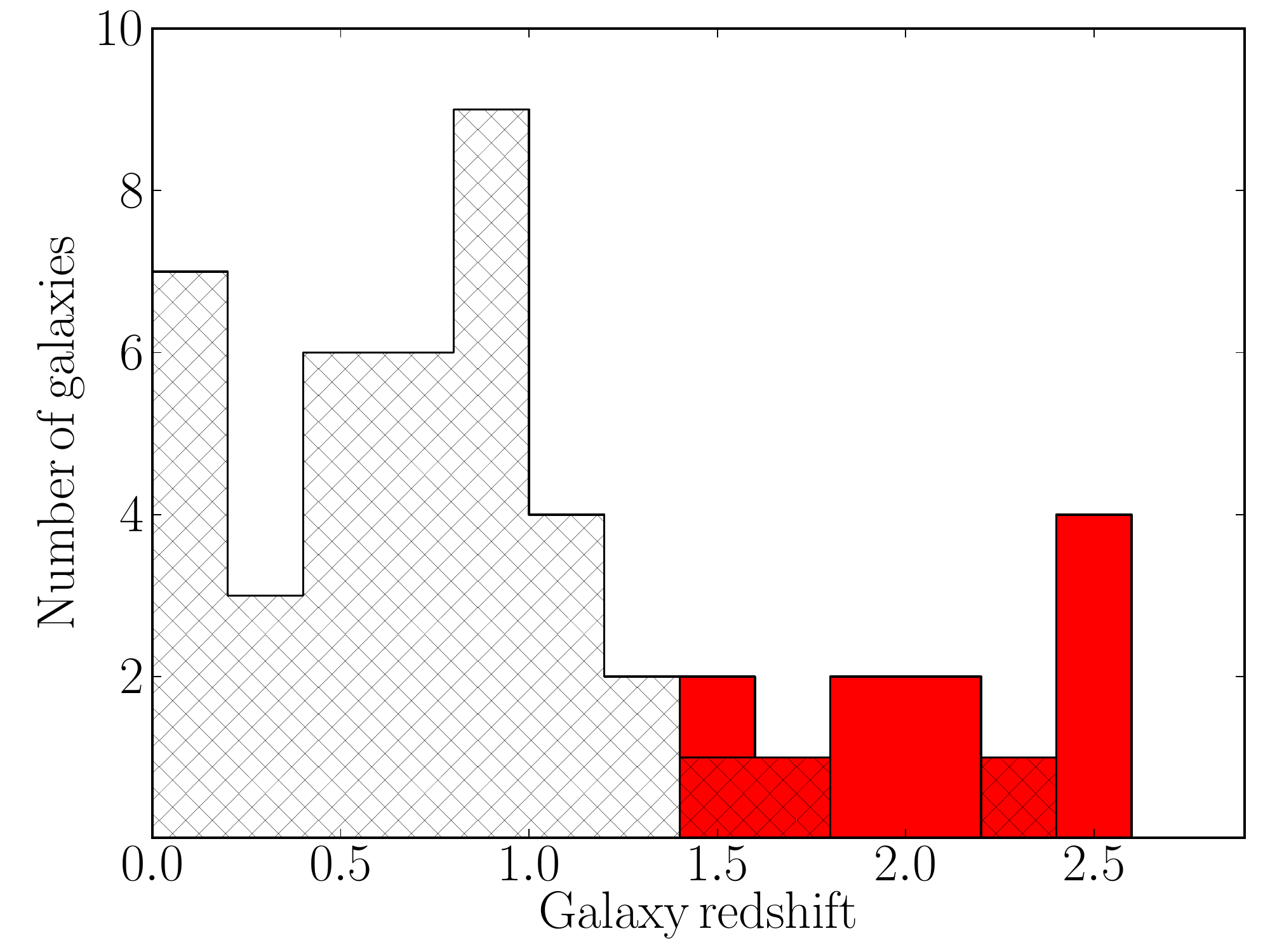}
\caption{Histogram of GRB host redshifts with published spectroscopy. Hatched data were taken from the literature \citep[][including updates from \texttt{http://www.grbhosts.org}]{2009ApJ...691..182S}, red from this work.}
\label{zhist}
\end{figure}

The median redshift is $\tilde{z}=2.1$ for all the 13 galaxies or $\tilde{z}=2.3$ for the nine TOUGH hosts (Table~\ref{tab:obs}) with accurate X-shooter emission line redshifts. Due to our initial selection and the unique NIR sensitivity of X-shooter, this is much higher than the median redshift of hosts ($\tilde{z}=0.7$) for which host spectroscopy has been published before the TOUGH survey (Figure~\ref{zhist}). Our targets were furthermore chosen to maximize the success rate of a redshift determination, and are thus skewed to the lower redshift end of the distribution of TOUGH GRBs with unknown redshift \citep{2012arXiv1205.3162H}. 

The median (mean) $R$-band brightness of the galaxies with successful redshift measurement is $\tilde{R}=24.6\,\rm{mag}$ (mean $\langle{R}\rangle\sim24.8\,\rm{mag}$). The galaxies for which we could not obtain redshifts with our X-shooter spectroscopy were either observed under unfavorable sky conditions (GRBs 060306 and 060805A, see Table~\ref{tab:obs}) or significantly fainter ($\tilde{R}=26.0\,\rm{mag}$, GRBs 050406, 060923A, 060923C, 070224) than the ones with redshift measurements. They are thus likely located at an even higher redshift than $\tilde{z}=2.1$ on average (see Section~\ref{seleff} and also \citealp{2012arXiv1205.3162H})

For GRB hosts above $z\gtrsim1.6$, the redshifted, resonant Ly$\alpha$ transition is within the spectral response of X-shooter. However, we do not detect significant Ly$\alpha$ emission in any of the 17 X-shooter spectra (after excluding those hosts where Ly$\alpha$ is securely out of the X-shooter band-pass). This is due to a combination of several effects: 13 out of 17 GRBs had no detection of an optical afterglow, and 10 out of the 17 were firmly dark (and thus dusty). Dust plays a crucial role in the escape fraction of Ly$\alpha$ photons, and the strength of the Ly$\alpha$ line \citep{2003A&A...406L..63F, 2012arXiv1205.3779M}. In addition also the insignificant depth of our observations (e.g., GRBs 060306 and 060805A) as well as galaxy faintness (e.g., GRBs 050406, 070224) contribute to the Ly$\alpha$ non-detection. 

{But even for hosts with a bright optical afterglow (indicative of little dust along the GRB sight line), vigorous star formation, and an appropriate redshift, such as the galaxies hosting GRBs 060604 and 070129, Ly$\alpha$ emission is not ubiquitous (see also Figure 11 of \citealp{2012arXiv1205.3779M}). This could be indicative of a heterogeneous distribution of dust in the inter-stellar medium of the hosts, as the afterglow probes only a specific sight-line, while the Ly$\alpha$ emission is an integrated property of the galaxy. A quantitative and more in depth discussion of the Ly$\alpha$ properties of our sample is beyond the scope of this paper and deferred to a future work. }

For the six GRBs~050406, 060306, 060805A, 060923A, 060923C and 070224 no emission lines were detected, and our observations hence do not provide an accurate redshift. {Based on the detection of the host galaxy in the FORS/$R$-band \citep[Figure~\ref{fct} and][]{Daniele2012}, the redshift of these six GRBs is constrained to $z < 5$. The detection of the afterglow of GRB~050406 in the UVOT/$u$-band limits its redshift further to $z \lesssim 3.2$ with a proposed photo-$z$ of $z\sim2.4$ by \citet{2006ApJ...643..276S}. Additional redshift constraints based on the detection of continuum emission in low-resolution host spectroscopy are $z \lesssim 2.8$ for GRB 060923A \citep{2008MNRAS.388.1743T} and $z \lesssim 2.5$ for GRB~060805A and GRB~060306 \citep{2012ApJ...752...62J}.}

\subsubsection{GRB 050819}
\label{050819}

A bright emission line is detected in the X-shooter spectrum of GRB~050819 at a wavelength of $17545\,\AA$, which could be interpreted as \oii~ at $z=3.71$, \oiii($\lambda5007$) at $z=2.504$ or \ha~at $z=1.673$. In the case of \oii, we would expect to resolve the doublet of \oii($\lambda\lambda\,3726, 3729)$ with two components of comparable intensity separated by $14\,\AA$. If the line was \ha, then \oiii($\lambda5007$) and/or \oii~should be detected for typical line ratios. Hence, we interpret the line as \oiii($\lambda$5007) at $z=2.504$. The marginal flux seen at the expected positions of \oii~and \hb~at $z=2.504$ with $\rm{S/N }\lesssim2$ (Figure~\ref{lines}) further supports our association. \ha~at $z=2.504$ is redshifted to $\lambda_{\rm obs}=22998\,\AA$ and our non-detection of \ha~is caused by the increased thermal background of X-shooter in this wavelength regime.

\subsubsection{GRB 060306}

A redshift of $z=3.5$ for GRB~060306 based on the detection of a single emission line interpreted as \oii~via X-shooter spectroscopy was reported by \citet{2011arXiv1112.1700S}. Our X-shooter data are too shallow to confirm or reject the line detection. Photometric and spectroscopic observations, however, indicate a lower redshift solution with $z \lesssim 2.5$ \citep{2012ApJ...752...62J}. We therefore consider an association of the line with \oiii($\lambda$5007) at $z=2.35$ or \ha~at $z=1.55$ more plausible. 

\subsubsection{GRB 060719}
\label{060719}
We detect a single, bright emission line at $16616\,\AA$ in the X-shooter spectrum of the dark GRB~060719 ($\beta_{\rm{OX}}<-0.1$). \oii~is readily excluded as the line is not resolved into two components. Using the redshift constraint ($z < 2.0$) from the host continuum emission \citep{2012ApJ...752...62J}, we interpret this line as \ha~at $z=1.532$. The flux increase seen at the corresponding wavelength of \oii~($\lambda_{\rm obs}=9440~\AA$) strongly supports this interpretation. \oiii($\lambda$5007) at $z=1.532$ is located at $\lambda_{\rm obs}=12677~\AA$, a region dominated by bright skylines. We hence consider the redshift robust.

\subsubsection{GRB 060814}

A redshift of $z=1.92$ for the dark GRB~060814 ($\beta_{\rm{OX}}<-0.1$) based on these data was reported by \citet{2011arXiv1112.1700S}. The system hosting GRB~060814 consists of two galaxies A and B separated by 1\farc{3} \citep[Figure 1 and][]{2012ApJ...752...62J} at redshifts of $z_{\rm A}=1.923$ and $z_{\rm B}=1.920$. The location of the NIR afterglow is consistent with galaxy A, and $z_{\rm A}=1.923$ is the redshift of GRB~060814. Galaxy C from Figure~\ref{fct} is an unrelated foreground object.

\subsubsection{GRBs 050915A, 051001, 060604, 070103, 070129 and 070419B}

These six hosts from the TOUGH sample have at least two emission lines detected at high significance (Figure~\ref{lines}). The redshift of the galaxies is hence uniquely determined, and the redshift of the GRB only relies on the association between burst and galaxy \citep{Daniele2012}. This association is very robust for GRBs~050915A, 060604, 070129 and 070419B because of an accurate position from an optical/NIR afterglow (Figure~\ref{fct}). Only for GRBs~051001 and 070103 is it based on the \textit{Swift}/XRT position with uncertainties of 1\farc{7} and 1\farc{5} at 90\% confidence (Figure~\ref{fct}). The corresponding redshifts are provided in Table~\ref{tab:obs} and Figure~\ref{lines}.

\subsubsection{GRB 071021}
The afterglow of the dark ($\beta_{\rm{OX}}\lesssim0.4$) GRB~071021 was detected as a fading source in the NIR only \citep{2007GCN..6968....1C}. The galaxy at a position consistent with the afterglow (Figure~\ref{fc}) has a redshift of $z=2.452$ based on the detection of the emission lines of \hb, \oiii,~and \ha~(Figure~\ref{lines}). Our spectroscopic redshift rules out a high-redshift nature for this event, and the red color of the afterglow is attributed to extinction by dust in the ISM of the host.

\subsubsection{GRB 080207}

The luminous and massive host of the dark ($\beta_{\rm{OX}}\lesssim0.3$) GRB~080207 was studied in detail photometrically by \citet{2011ApJ...736L..36H} and \citet{2011arXiv1109.3167S}, yielding photometric redshifts of $z^{\rm{phot}}\sim2$. The spectroscopic redshift based on \ha~and \oiii~is $z^{\rm{spec}}=2.086$, and thus confirms the estimation by the preceding authors. 

\subsubsection{GRB 090113}

The optical afterglow of the dark GRB~090113 was not detected in NOT $R$-band imaging \citep{2009GCN..8810....1D}, which yields  $\beta_{\rm{OX}}\lesssim0.3$. The galaxy associated with the burst via a NIR afterglow position (Figure~\ref{fc}, A.~J. Levan, priv. communication) has a redshift of $z=1.749$ based on the detection of the emission lines of \oii~ and \ha~(Figure~\ref{lines}). At this redshift, \hb~and \oiii~are located in regions of limited atmospheric transmission.

\subsubsection{GRB 090407}
The afterglow of the dark GRB~090407 was not detected in deep VLT/FORS or GROND optical/NIR imaging \citep{2009GCN..9108....1M, 2009GCN..9109....1A}, implying $\beta_{\rm{OX}}\lesssim0.4$. The galaxy associated with GRB~090407 via \textit{Chandra} observations (Figure~\ref{fc}, A.~J. Levan, priv. communication) has a redshift of $z=1.448$, which relies on the detection of \oii~and \ha~(Figure~\ref{lines}).  At this redshift, \hb~and \oiii~are located in a region dominated by bright sky-lines.

\subsection{Absorption Properties of dark GRBs}
\label{darkabs}
Having an accurate redshift measurement enables quantitative studies of the material along the line of sight. Low-energy X-ray afterglow photons are, for example, absorbed by intervening metals. This metal column density is typically converted into an equivalent hydrogen column at the redshift of the GRB assuming solar metallicity $N_{\rm{H,X}}$. The inferred $N_{\rm{H,X}}$ values are generally much higher than the neutral hydrogen column density measured from GRB DLAs, and the exact physical nature and location of the material causing the soft X-ray absorption excess is currently not fully understood \citep[e.g.,][]{2011A&A...525A.113S, 2012ApJ...754...89W}. 

\begin{figure}
\includegraphics[angle=0, width=0.99\columnwidth]{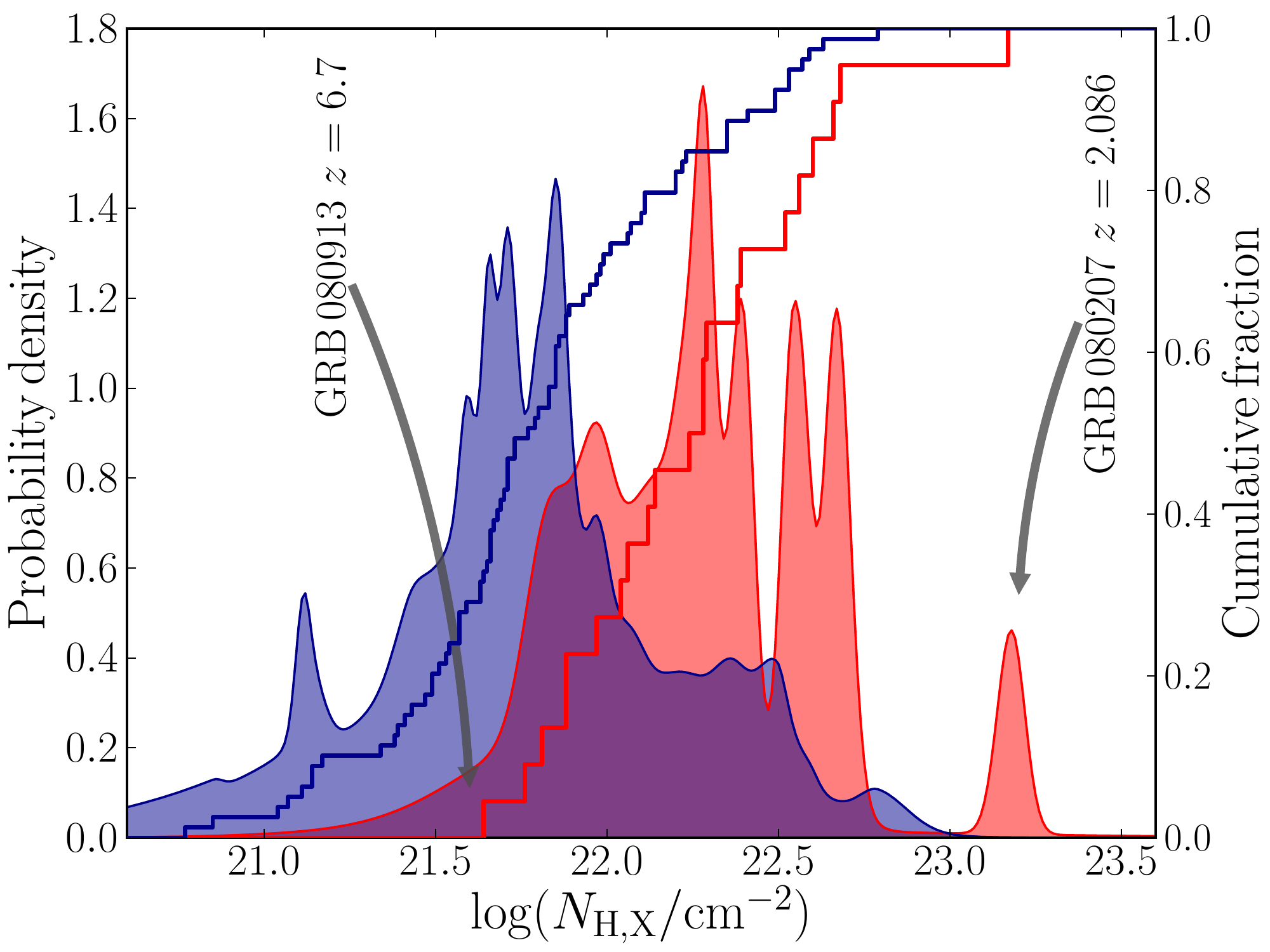}
\caption{Normalized probability density function and cumulative distribution of soft X-ray absorption at the burst's redshift for 79 optically-bright \textit{Swift} GRBs in blue \citep{2010MNRAS.402.2429C} and 22 dark GRBs in red with spectroscopic redshifts (Table~\ref{tab:nhd}). The probability density function (PDF) is derived by assuming continuous normal distributions of $N_{\rm{H,X}}$ for single GRBs with asymmetric width set by the measured uncertainties \citep[see e.g.,][]{2009MNRAS.397.1177E}. The plotted PDF is then the sum over the individual distributions of $N_{\rm{H,X}}$ normalized to the number of GRBs. The location of the high-redshift, dark GRB 080913 \citep{2009ApJ...693.1610G} and the dusty, dark GRB~080207 are indicated in the plot.}
\label{nhhist}
\end{figure}

Figure~\ref{nhhist} shows the probability density function and cumulative distribution of $N_{\rm{H,X}}$ at the burst's redshift for dark GRBs with $\beta_{\rm{OX}}<0.5$ and an accurate redshift. We selected the dark GRBs from this work as well as several events from the literature \citep{2011A&A...526A..30G, 2011arXiv1108.0674K, Dan2012}. X-ray data for all events were taken from the \textit{Swift}-XRT online repository \citep{2007A&A...469..379E, 2009MNRAS.397.1177E} {and analyzed in standard manner in \texttt{XSPEC12} using abundances from \citet{1989GeCoA..53..197A}. The selected bursts, their redshifts, and $N_{\rm{H,X}}$ values are given in Table~\ref{tab:nhd}. As a comparison sample, we used 79 optically-bright, redshift-selected \textit{Swift} GRBs and their $N_{\rm{H,X}}$ values up to June 2009 from Table 1 in \citet{2010MNRAS.402.2429C}. }

\begin{deluxetable}{ccccc}
\tabletypesize{\scriptsize}
\tablecaption{$N_{\rm{H,X}}$ Values for the Dark GRBs With Spectroscopic Redshifts
\label{tab:nhd}}
\tablecolumns{5} 
\tablewidth{0pt}
\tablehead{
	\colhead {GRB} & 
	\colhead {Redshift} & 
	\colhead {$\beta_{\rm{OX}}$} & 
	\colhead {$N_{\rm{H,X}}$} & 
	\colhead {References} \\
	 {} & 
	 {} & 
	 {} & 
	 {($10^{22}\,\rm{cm}^{-2}$)} & 
	 {}} 
\startdata
050915A  & 2.53 & $<0.4$ & $1.1^{+0.7}_{-0.6}$ & (1, 2) \\
060210   & 3.91 & $0.4$ & $2.4\pm0.4$ & (2)  \\
060719  & 1.53 & $<-0.1$ & $1.9^{+0.5}_{-0.4}$ & (1, 2) \\
060814  & 1.92 & $<-0.1$ & $3.4^{+0.3}_{-0.4}$ & (1, 2) \\
061222A   & 2.09 &$<0.2$ & $3.7\pm0.4$ & (2) \\
070103   & 2.62 & $<0.5$ & $4.1^{+1.3}_{-1.1}$ & (1) \\
070306   & 1.50 & $<0.2$ & $2.5\pm0.3$ &  (2) \\
070419B   & 1.96 & $0.3$ & $0.8\pm0.2$ & (1) \\
070521   & 1.35 & $<-0.1$ & $4.8\pm0.7$ & (2) \\
070802   & 2.45 & $0.5$ &$0.8^{+0.8}_{-0.6}$ & (2, 3) \\
071021   & 2.45 & $<0.4$& $1.3\pm0.4$ & (1) \\
080207   & 2.09 & $<0.3$ & $15.1^{+2.3}_{-2.2}$ & (1) \\
080805   & 1.50 & $0.4$ &$1.2^{+0.7}_{-0.5}$ & (2, 3)  \\
080913   & 6.7 & $<0.1$ &$0.4^{+6.2}_{-0.4}$ & (2, 3) \\
081109   & 0.98 & $<0.4$ &$0.9\pm0.2$ & (3, 4) \\
081221   & 2.26 & $<0.2$ &$4.6\pm0.6$ & (5) \\
090102   & 1.55  & $0.5$ &$0.7^{+0.2}_{-0.1}$ & (3, 6) \\
090113   & 1.75 & $<0.3$ & $2.0^{+0.8}_{-0.6}$& (1) \\
090407   & 1.45 & $<0.4$ & $1.8\pm0.3$ & (1) \\
090812   & 2.45 & $0.5$ & $0.6\pm0.4$&  (3, 7) \\
090926B   & 1.24 & $0.4$ & $1.4^{+0.5}_{-0.4}$& (3, 8) \\
100621A   & 0.54 & $0.4$ & $1.9\pm0.2$ & (4, 9) \\
\enddata
\tablecomments{References: (1) This work, (2) \citet{2009ApJS..185..526F}, (3) \citet{2011A&A...526A..30G}, (4) \citet{2011arXiv1108.0674K}, (5) \citet{2011arXiv1112.1700S}, (6) \citet{2009GCN..8766....1D}, (7)  \citet{2009GCN..9771....1D}, (8) \citet{2009GCN..9947....1F}, (9) \citet{2010GCN.10876....1M}  }
\end{deluxetable}

The median (mean) X-ray column densities $\log(N_{\rm{H,X}})$ for both distributions are $\log(N_{\rm{H,X}}/\rm{cm}^{-2})=21.8\,(21.8)$ for the optically bright and $\log(N_{\rm{H,X}}/\rm{cm}^{-2})=22.3\,(22.2)$ for the dark \textit{Swift} events. Clearly, dark (which are predominantly dusty) bursts have significantly higher $N_{\rm{H,X}}$ values with its most extreme example GRB~080207 exceeding $N_{\rm{H,X}}>10^{23}\,\rm{cm}^{-2}$ (Figure~\ref{nhhist}). The K.-S.-test $p$-value is $2.3\times10^{-4}$ and the null-hypothesis, that both distributions are drawn from the same parent distribution, can hence be rejected at $3.7\,\sigma$ confidence in a normal probability distribution. 

This result is not unexpected. Per selection, we have compared dark bursts which are predominantly dusty (with a few high-redshift interlopers) against dust-poor sight-lines, and metals are a prerequisite of forming dust. There is hence a tight connection between the darkness of the afterglow and the soft X-ray absorption. It furthermore illustrates how our understanding of GRB and host galaxy properties is limited by the effect of incompleteness in redshift samples (see also \citealp{2009ApJS..185..526F, 2012MNRAS.421.1697C}), and how X-shooter spectroscopy of GRB hosts is capable of efficiently removing this bias.

\section{Conclusions}

We have presented optical/NIR spectroscopy of a sample of 19 \textit{Swift} GRBs primarily taken from the TOUGH survey. In total, we measure 13 redshifts, including the redshifts of nine dark GRBs. The median redshift of these 13 galaxies is $\tilde{z}=2.1$, and it is $\tilde{z}=2.3$ for our nine bursts from the TOUGH survey. This establishes NIR host spectroscopy as an efficient tool for redshift determination in the redshift desert at $z \gtrsim 1.4$ (Figure~\ref{zhist}) and in those cases where optical/NIR afterglow absorption line spectroscopy is not successful. 

X-shooter's sensitivity, resolution and wavelength coverage are ideally suited to perform redshift measurements for star-forming galaxies such as GRB hosts with brightnesses $R\sim24.5\,\rm{mag}$ and modest exposure times. Our observations yield redshifts for even fainter targets (GRBs~080207 with  $R=25.8\pm0.2\,\rm{mag}$ or 090407 with $\rm{F606W}>27\,\rm{mag}$) in cases when the galaxies are NIR bright ($K\lesssim 21\,\rm{mag}$) and vigorously star forming.

In particular for highly-extinguished GRBs and afterglows, NIR host spectroscopy provides means to establish an accurate redshift, and thus lifts the bias in previous GRB samples, which typically lack redshift determinations for dark bursts.  

This is a further step towards complete GRB samples, and hence a better understanding of the conditions in which GRBs form. A census of GRB hosts without physical selection biases is the key for measuring the fraction of star-formation that is traced by long GRBs. Samples like TOUGH are therefore of fundamental importance when using GRBs as tools to investigate galaxy evolution and star-formation in the early Universe.

\acknowledgements

T.K. acknowledges support from a Marie-Curie Intra-European Fellowship. D.M. acknowledges support from the Instrument Center for Danish Astrophysics.  B.M.-J. and J.P.U.F acknowledge support from the ERC-StG grant EGGS-278202. P.J. acknowledges support by a Project Grant from the Icelandic Research Fund. The Dark Cosmology Centre is funded by the Danish National Research Foundation. We thank the referee for very valuable and constructive comments, G. Tagliaferri for imaging data of GRB~071021, G. Leloudas for comments on the manuscript, P. Vreeswijk for support and D. Perley for sharing his results before publication. This research has made use of the GHostS database (\texttt{http://www.grbhosts.org}), which is partly funded by Spitzer/NASA grant RSA Agreement No. 1287913. This work made use of data supplied by the UK \textit{Swift} Science Data Centre at the University of Leicester. 

{\it Facilities:} \facility{VLT:Kueyen (X-shooter)}, \facility{VLT:Antu (FORS2)}, \facility{VLT:Melipal (ISAAC)}, \facility{HST (WFC3)}


\end{document}